\pdfoutput=1

\documentclass[aps,twocolumn,amsmath,amssymb,preprintnumbers,floatfix,prl,superscriptaddress,longbibliography]{revtex4-2}
\usepackage[utf8]{inputenc}
\usepackage{eucal}
\usepackage{bm}
\usepackage{siunitx}
\usepackage{comment}
\usepackage{cancel}
\usepackage{enumerate}
\usepackage{amsfonts}
\usepackage{amsmath}
\usepackage{amssymb}
\usepackage{color}
\usepackage{xcolor}
\usepackage{soul}

\usepackage{graphicx}
\usepackage[caption=false]{subfig}
\graphicspath{{plots/}}
\usepackage{tikz}
\usetikzlibrary{calc,shapes}
\tikzset{
odot/.style={
  circle,
  inner sep=0pt,
  node contents={$\odot$},
  scale=1.5
},
circ/.style={
  circle,
  draw,
  thick,
  minimum size=3mm,
  inner sep=0
},
odot2/.style={
  circ,
  path picture={\fill circle[radius=1.6pt];}
}
}

\usepackage[colorlinks,allcolors=blue]{hyperref}
\usepackage[capitalize]{cleveref}
\usepackage{cleveref}

\definecolor{DarkGray}{rgb}{0.7,0.7,0.7}
\definecolor{DarkRed}{rgb}{0.65,0,0}%
\definecolor{Green}{rgb}{0,0.3,0.3}
\definecolor{Purple}{rgb}{0.3,0,0.65}
\definecolor{Red}{rgb}{1,0,0}
\definecolor{Blue}{rgb}{0,0,0.85}
\definecolor{Magenta}{rgb}{1,0,1}

\newcommand{\Real}{{\mathrm{Re}}}   

\DeclareMathOperator{\diag}{diag} 

\newcommand{\Eq}{Eq.~}
\newcommand{\Eqs}{Eqs.~}

\newcommand{\ie}{\textit{i.e. }}

\newcommand{\be}{\begin{equation}}
\newcommand{\ee}{\end{equation}}
\newcommand{\bs}{\begin{split}}
\newcommand{\es}{\end{split}}
\newcommand{\bse}{\begin{subequations}}
\newcommand{\ese}{\end{subequations}}
\newcommand{\bal}{\begin{align}}
\newcommand{\eal}{\end{align}}

\newcommand{\prlsection}[1]{\textit{#1}.\kern0.05em---\kern0.05em\ignorespaces}

\begin{document}
\title{Curvature-induced long ranged supercurrents in diffusive SFS Josephson Junctions, with dynamic $0-\pi$ transition}
\author{Tancredi Salamone}
\email[Corresponding author: ]{tancredi.salamone@ntnu.no}
\affiliation{Center for Quantum Spintronics, Department of Physics, NTNU Norwegian University of Science and Technology, NO-7491 Trondheim, Norway}
\author{Mathias B.M. Svendsen}
\affiliation{Center for Quantum Spintronics, Department of Physics, NTNU Norwegian University of Science and Technology, NO-7491 Trondheim, Norway}
\author{Morten Amundsen}
\affiliation{Nordita, KTH Royal Institute of Technology and Stockholm University,
Hannes Alfvéns väg 12, SE-106 91 Stockholm, Sweden}
\author{Sol Jacobsen}
\affiliation{Center for Quantum Spintronics, Department of Physics, NTNU Norwegian University of Science and Technology, NO-7491 Trondheim, Norway}

\begin{abstract}
We report that spin supercurrents can be induced in diffusive SFS Josephson junctions without any magnetic misalignment or intrinsic spin orbit coupling. Instead, the pathway to spin triplet generation is provided via geometric curvature, and results in a long ranged Josephson effect. In addition, the curvature is shown to induce a dynamically tunable $0-\pi$ transition in the junction. We provide the analytic framework and discuss potential experimental and innovation implications.
\end{abstract}
\maketitle


\textit{Introduction}.--- 
In the last two decades there have been substantial advances in the experimental realization of curved nanostructures. Since the realization of nanotubes by rolling up thin solid films \cite{schmidt2001}, many new techniques of bending, wrinkling and buckling nanostructures in up to three dimensions have been developed \cite{cendula2009,Xu154}, as well as direct growth on curved templates \cite{Das2019}, electronbeam lithography \cite{Lewis2009,Burn2014,Volkov2019} and many more (see e.g. \cite{Streubel2016} and references therein). These techniques open a broad new range of spintronic device design, and have already been shown to enable independent control of spin and charge resistances \cite{Das2019}.

Geometric curvature introduces two main effects: a quantum geometric potential, producing many interesting phenomena at the nanoscale \cite{Cantele2000,Aoki2001,Encinosa2003,Ortix2010}, and a strain field leading to a curvature-induced Rashba spin-orbit coupling (SOC), with strength proportional to the curvature \cite{gentile2013}. Several studies have investigated new properties triggered by curvature, e.g. in semiconductors \cite{Nagasawa2013,Gentile2015,Ying2016,Chang2017,Francica2019}, magnets \cite{Streubel2016,Volkov2019,Das2019} and superconductors \cite{Turner2010,Francica2020,Chou2021}. 
Curved nanostructures with induced superconductivity can display geometric control of spin-triplet correlations in the clean limit \cite{Ying2017}, and proximizing a superconductor with a curved semiconductor can result in topological edge states \cite{Gentile2015}. The curved topological superconductor/straight semiconductor Josephson junction counterpart has been predicted to display a $0-\pi$ transition and $\phi$-junction behaviour \cite{Francica2020}. 

\begin{figure}[b]
    \centering
    \includegraphics[width=0.85\columnwidth]{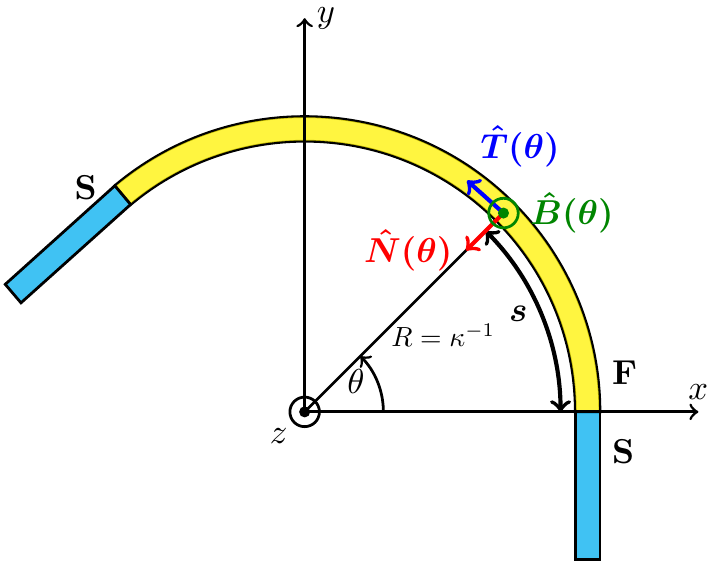}
    \caption{Model system. SFS junction having a ferromagnet with constant curvature as weak link. The three orthonormal unit vectors $\hat{T}(\theta)$, $\hat{N}(\theta)$ and $\hat{B}(\theta)$ identifying the curvilinear coordinates are also shown.}
    \label{fig:illustration}
\end{figure}

Hybrid structures of superconductors and ferromagnets are of great interest for the field of superconducting spintronics \cite{linder_nphys_15,Eschrig2011} since at the superconductor/ferromagnet (SF) interface the proximity effect allows the property of one material to ``leak'' into the other \cite{buzdin_rmp_05,bergeret_rmp_05,Lyuksyutov2005}. A coexistence of superconductivity and magnetism may therefore enable data processing, encoded in spin and charge degrees of freedom, to be performed without the heat loss associated with traditional electronics. In diffusive heterostructures, which cover a range of commonly available materials that may have impurities or sub-optimal interface transparencies, conventional $s$-wave superconducting correlations typically penetrate a ferromagnet for extremely short distances, proportional to $\sqrt{D_F/h}$, with $D_F$ the diffusion constant and $h$ the exchange field strength. Significant theoretical and experimental effort has focused on the conversion of singlet correlations into so-called long range triplet correlations (LRTC), which penetrate for longer distances, on the order of $\sqrt{D_F/T}$, where $T$ is the temperature. This conversion can take place in the presence of magnetic inhomogeneities \cite{bergeret_prl_01,khaire,robinson} or due to intrinsic spin-orbit coupling either in the superconductor or in the ferromagnet \cite{BergeretTokatly2013,BergeretTokatly2014}. The role of geometric curvature as a source of designable and dynamically alterable SOC in diffusive structures has not been investigated in this context, and we address this here. By considering a model SFS junction with constant curvature shown in Fig.~\ref{fig:illustration}, we show that the curvature alone can induce long ranged supercurrents due to the generation of triplet correlations. Moreover, we show that these systems display a tunable $0-\pi$ transition.

The possibility of $0-\pi$ state switching has been of much interest, in part due to its potential role in solid state quantum computing \cite{Buzdin1982,Buzdin2003,buzdin_rmp_05,Ryazanov2001,Kontos2002,Mooij1999}.
Investigations have confirmed the transition can be governed by altering the length of the ferromagnetic weak link. However, this is not practicable to do in-situ and must be done by preparing multiple samples of different lengths. It has recently been predicted that the $0-\pi$ transition can also be accessed out of equilibrium, by altering the strength of the SOC via voltage gating \cite{Bujnowski2019}. In this Letter, we show that dynamically changing the curvature of the magnet via in-situ strain manipulation, for example via photostriction, piezoelectrics or thermoelectric effects \cite{Kundys2015,Matzen2019}, allows for a single-sample $0-\pi$ transition in the diffusive regime, without the need to apply a varying voltage. Moreover, we show that curvature can yield long-ranged Josephson currents without any magnetic inhomogeneities or intrinsic SOC.

\vspace{3mm}

\textit{Theoretical framework}.---
A fundamental tool for the study of curved structures is the thin-wall quantization procedure, where the quantum motion of a particle in a 2D curved surface is treated as equivalent to the motion in a 3D space with the addition of lateral quantum confinement \cite{jensen1971, daCosta1981}. This procedure allows one to derive the Hamiltonian for the motion of electrons constrained to a curved planar one-dimensional structure \cite{Ortix2011,Ortix2015}. 

When dealing with a ferromagnet, further effects of the curvature must be taken into account, namely in how it affects its exchange field. Recent studies have developed a fully 3D approach for thin magnetic shells of arbitrary shape and extended it to 2D shells and 1D wires \cite{gaididei2014,Sheka_2015}. This showed that curvature induces two effective magnetic interactions: an effective magnetic anisotropy and an effective Dzyaloshinskii-Moriya interaction (DMI). When dealing with a 1D curved wire below a certain critical curvature, the magnetic anisotropy and the DMI, which results in an effective Rashba SOC, both combine to give an effective field tangential to the wire. The geometrically defined SOC is therefore both designable and tunable, and gives greater freedom in the manipulation of superconducting proximity effects.

We parametrize the curve by its arc length $s$, and define a set of three orthonormal unit vectors $\hat{T}(s)$, $\hat{N}(s)$, $\hat{B}(s)$ representing the tangential, normal and binormal curvilinear coordinates respectively, as indicated in Fig.~\ref{fig:illustration}. These obey the following Frenet-Serret-type equation of motion:

\be
\begin{pmatrix} \partial_s\hat{T}(s) \\ \partial_s\hat{N}(s) \\ \partial_s\hat{B}(s) \end{pmatrix} = \begin{pmatrix} 0 & \kappa(s) & 0 \\ -\kappa(s) & 0 & 0 \\ 0 & 0 & 0 \end{pmatrix}\begin{pmatrix} \hat{T}(s) \\ \hat{N}(s) \\ \hat{B}(s) \end{pmatrix},
\label{FStype}
\ee

\noindent where $\kappa(s)$ is the curvature of the wire, whose role and effect will be discussed in detail below. Deriving the Hamiltonian for a wire, which may include intrinsic SOC in general, we find \cite{Ortix2015}:

\be\begin{split}
H =& -\frac{\hbar^2}{2m}\partial_s^2-\frac{\hbar^2}{8m}\kappa(s)^2-i\hbar\alpha_N\sigma_B\partial_s\\&+i\hbar\alpha_B\left(\sigma_N\partial_s-\frac{\kappa(s)}{2}\sigma_T\right).
\label{H1D}
\end{split}\ee

\noindent The SOC constants $\alpha_{N,B}$ represent the spin-orbit field with axis along the normal and binormal direction respectively, and $\sigma_{T,N,B}(s)=\bm\sigma\cdot\{\hat T,\hat N, \hat B\}(s)$ are the set of three Pauli matrices in curvilinear coordinates. By using \Eqs\eqref{FStype} we can incorporate the last three terms in \Eq\eqref{H1D} in a SU(2) spin-orbit field term: 

\be
\bm A=(\alpha_N\sigma_B-\alpha_B\sigma_N,0,0),
\label{SOfield}
\ee

\noindent which has a vector structure in the geometric space and a $2\times2$ matrix structure in spin space. It is worth distinguishing between the two terms entering the SU(2) field, namely $\alpha_B$ and $\alpha_N$. The former represents the intrinsic, not induced by the curvature, SOC term which may or may not exist according to the material taken into consideration. The latter is curvature-induced, and is proportional to the curvature strength. In natural units we have $\alpha_N=g\lambda\kappa(s)/(4m)$, where $g$ is the g-factor and the parameter $\lambda>0$ is a characteristic energy scale for the material. Inspection of the relevant diffusion equations for the system shows that $\alpha_N$ and $\kappa(s)$ appear together in such a way that the former always acts as a strengthening factor for the latter. Therefore, considering a material with no intrinsic term we can ignore spin-orbit coupling as a whole, and consider the $\kappa(s)$ term only.

Having set up the Hamiltonian, we employ Green functions in the diffusive limit at equilibrium. Here the dynamics are describable by the second-order partial differential Usadel equation \cite{Usadel1970}, which, with suitable boundary conditions, describes the diffusion of superconducting correlations inside the ferromagnet. Treating the case of diffusive equilibrium, it is sufficient to consider just the retarded component $\hat{g}_R$ of the quasiclassical Green function to describe the system \cite{belzig_review_98}.
Using \Eq\eqref{H1D} the Usadel equation in a curved ferromagnet with constant curvature reads (from now on we set $\hbar=1$):

\be
	D_F\partial_s\left(\hat{g}_R\partial_s\hat{g}_R\right)+i\left[\varepsilon\hat\tau_3+\hat{M},\hat{g}_R\right]=0,
	\label{usadel}
\ee

\noindent with $\hat\tau_3=\diag(1,1,-1,-1)$, $\varepsilon$ the quasiparticle energy and magnetization $\hat{M}=\bm h\cdot\diag(\bm\sigma,\bm\sigma^*)$. The components of both vectors $\bm h=(h_T,h_N,h_B)$ and $\bm\sigma=(\sigma_T,\sigma_N,\sigma_B)$ are expressed in curvilinear coordinates. To solve the Usadel equation we employ the Kuprianov-Lukichev boundary conditions \cite{KuprianovLukichev1988}:

\be
	L_j\zeta_j\hat{g}_{Rj}\nabla_{I}\hat{g}_{Rj}=\left[\hat{g}_{R1},\hat{g}_{R2}\right].
\ee

\noindent Here $\nabla_{I}$ is the derivative at the interface, $j$ refers to the various components of the hybrid system, with $j=1,2$ denoting the materials on the left and right side of the relevant interface, $L_j$ represents the length of the material and $\zeta_j=R_B/R_j$ is the interface parameter given by the ratio between the barrier resistance $R_B$ and its bulk resistance $R_j$.

If desirable, the intrinsic SOC can be retained, in which case one also introduces the gauge covariant derivative \cite{BergeretTokatly2014}:

\be
	\partial_s(\,\cdot\,)\rightarrow\widetilde{\partial_s}(\,\cdot\,)\equiv\partial_s(\,\cdot\,)-i\left[\hat{A}_T,\cdot\,\right],
\ee

\noindent with $\hat{A}_T=\diag(A_T,-A_T^*)$ and $A_T$  is the tangential component of the SO field of \Eq\eqref{SOfield}.

To treat the system we will use the Riccati parametrization \cite{SchopohlMaki1995,JacobsenLinder2015} for the quasiclassical Green function:

\be
	\hat{g}_R=\begin{pmatrix} N(1+\gamma\tilde{\gamma}) & 2N\gamma \\ -2\tilde{N}\tilde{\gamma} & -\tilde{N}(1+\tilde{\gamma}\gamma) \end{pmatrix},
\ee

\noindent where the normalization matrices are $N=(1-\gamma\tilde{\gamma})^{-1}$ and $\tilde{N}=(1-\tilde{\gamma}\gamma)^{-1}$ and the tilde operation denotes $\tilde\gamma(\varepsilon)=\gamma^*(-\varepsilon)$. The Usadel equation \eqref{usadel} thus becomes:

\be\begin{split}
	D_F\left\{\partial^2_s\gamma+\right.&\left.2(\partial_s\gamma)\tilde{N}\tilde{\gamma}(\partial_s\gamma)\right\}=\\&-2i\varepsilon\gamma-i\bm{h}\cdot(\bm{\sigma}(s)\gamma-\gamma\bm{\sigma}^*(s)).
	\label{usadel_riccati}
\end{split}\ee

\noindent Here the dependence on the curvature is implicitly contained in the Pauli matrices $\sigma_{T,N,B}(s)$. 

We will consider our one-dimensional curved wire to be lying in the $xy$ plane as represented in Fig.~\ref{fig:illustration}, so that the set of three unit vectors is:

\begin{subequations}\begin{align}
 \hat T(s) &= - \sin\theta(s)\hat x + \cos\theta(s)\hat y,\\
 \hat N(s) &= -\cos\theta(s)\hat x - \sin\theta(s)\hat y,\\
 \hat B(s) &\equiv \hat{z},
\end{align}\end{subequations}

\noindent with $\theta(s) = \kappa s$. It is useful to note that, when considering \Eq\eqref{usadel_riccati}, the curved ferromagnet can be regarded as equivalent to a straight wire with a rotating exchange field, \ie a tangential exchange field in a curved wire is equivalent to a position dependent exchange field in a straight wire, varying as $\vec{h}(s) = h_0(\sin\theta(s),-\cos\theta(s),0)$, with $\theta(s) = \pi s/L_F$ and $L_F$ being the length of the ferromagnet.

\vspace{3mm}

\textit{Results}.---
Solving the Usadel equation, and therefore finding the quasiclassical Green function of the system, allows us to calculate many interesting quantities. In this work we will focus mainly on the charge current given by:

\be
	\frac{I_{Q}}{I_{Q0}}=\int_{-\infty}^{+\infty}\!\! d\varepsilon\mathrm{Tr}\left\{\hat\tau_3\left(\hat g_R\partial_s\hat g_R-\hat g_A\partial_s\hat g_A\right)\right\}\tanh(\beta\varepsilon/2).
 \label{Q_current}
\ee

\noindent Here $\hat g_A = -\hat\tau_3\hat g_R^\dagger\hat\tau_3$ is the advanced quasiclassical Green function and $\beta=(k_BT)^{-1}$ is the inverse temperature, with $k_B$ being the Boltzmann constant. Moreover, $I_{Q0}=N_0eD_FA\Delta_0/4L_F$, where $N_0$ is the density of states at the Fermi energy, $A$ the interfacial contact area and $\Delta_0$ the bulk gap of the two superconductors. Lengths and energies have been normalized to $L_F$ (which in turn is scaled with the superconducting coherence length $\xi_S$) and superconducting bulk gap $\Delta_0$ respectively, so that the integral on the right side of \Eq\eqref{Q_current} is dimensionless.

We investigate the system portrayed in Fig.~\ref{fig:illustration} by solving numerically \Eq\eqref{usadel_riccati} for various lengths $L_F$ of the ferromagnet and multiple curvatures $\kappa$ for each length. We set the interface parameter with both superconductors to be $\zeta=3$ and the temperature to $T=0.005T_c$. We consider the exchange field inside the curved ferromagnet to be tangential to its curvature profile at each point, $\bm h(s)\parallel\hat T(s)$, which we expect to be the case in 1D curved structures below a certain critical curvature \cite{Sheka_2015}.

\begin{figure}
         \centering
         \includegraphics[width=\columnwidth]{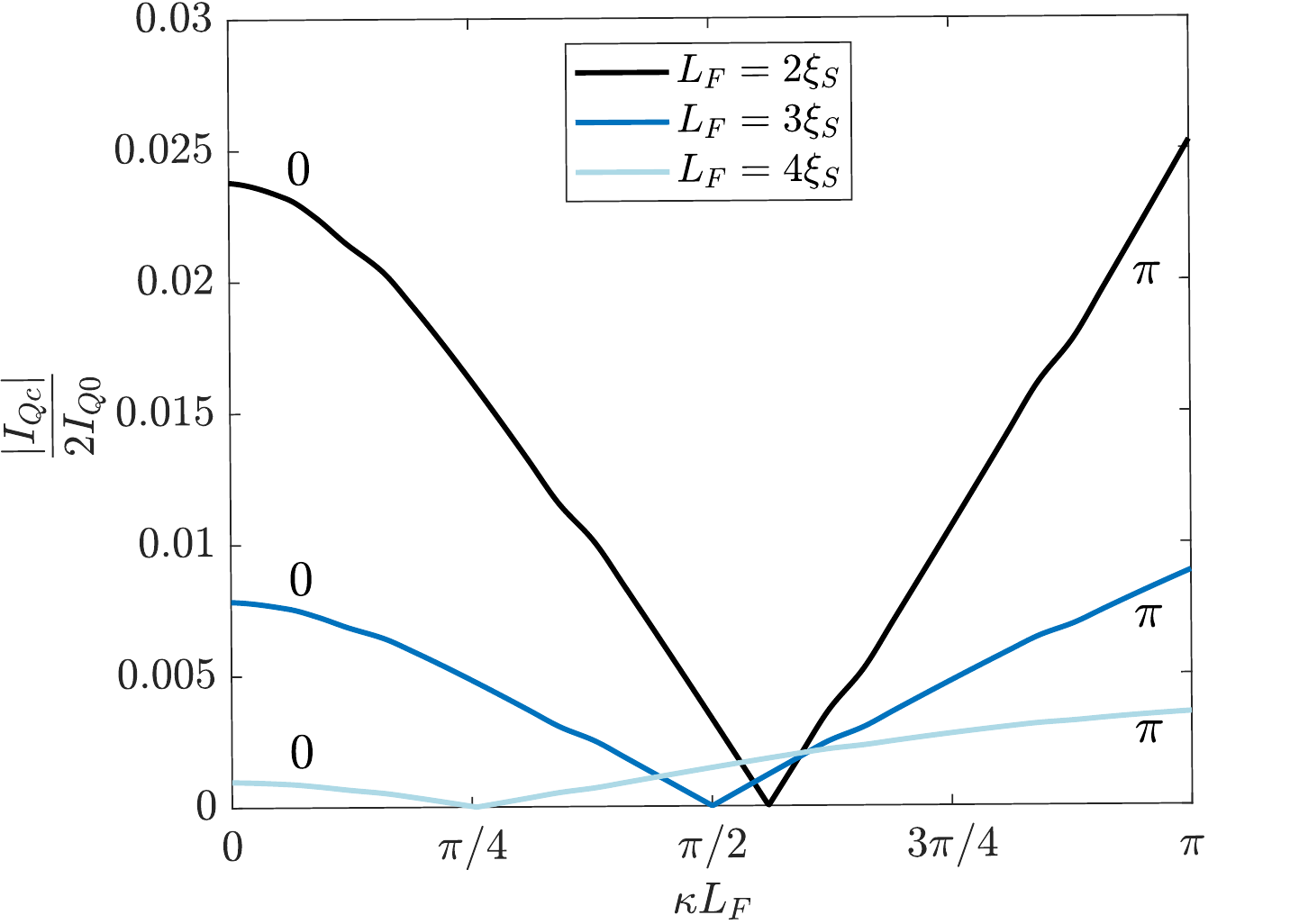}
         \caption{Magnitude of the critical current as a function of the curvature for different lengths $L_F$ of the ferromagnet, with $T=0.005T_c$, $\vec h = \Delta_0\hat{T}$, $\zeta=3$. A $0-\pi$ transition occurs when changing the curvature of the wire.}
         \label{IQc_L2}
\end{figure}

\begin{figure}
	\centering
	\includegraphics[clip,width=\columnwidth]{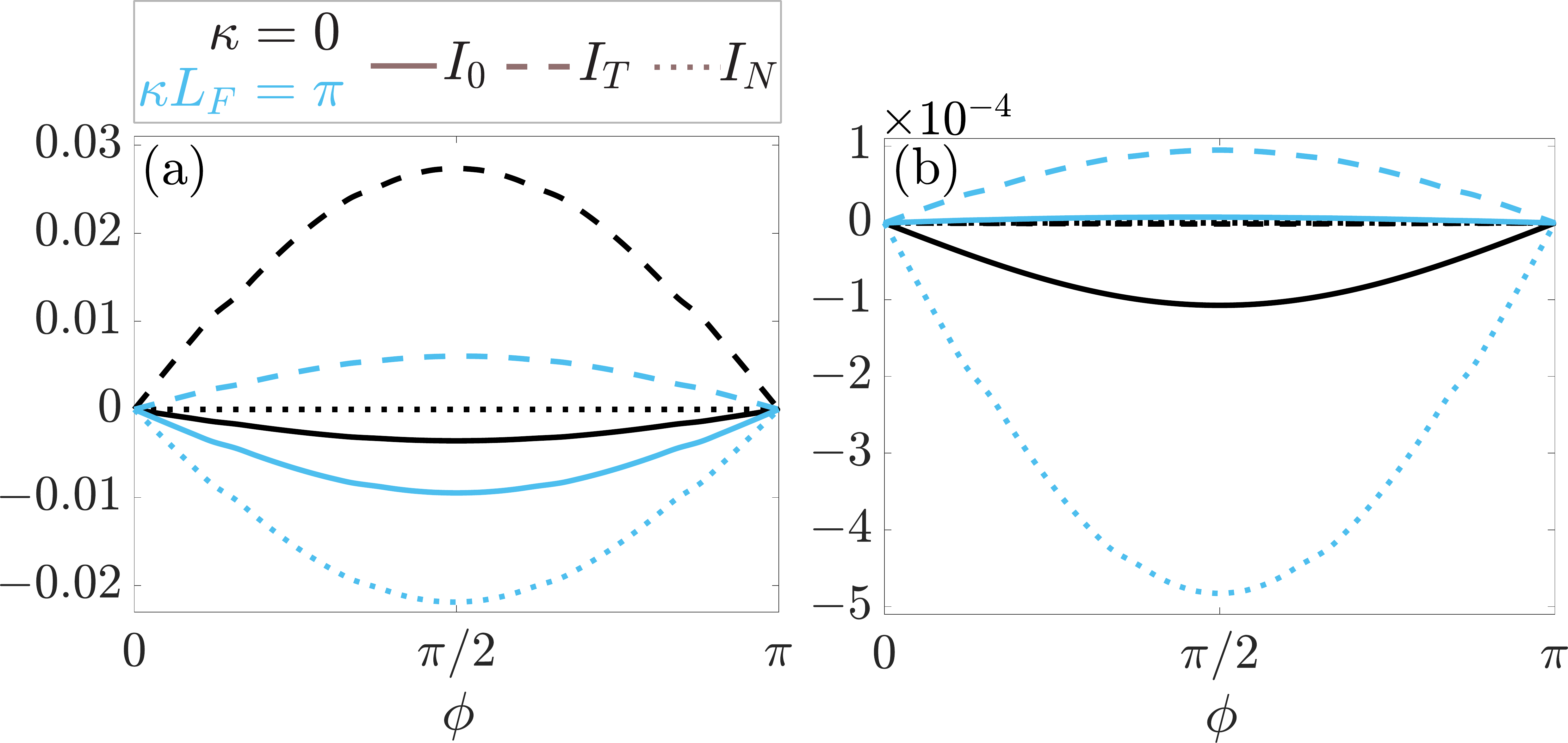}
	\caption{Charge current as a function of the phase difference $\phi$, showing separately the singlet (solid lines) and triplet (dashed lines) contributions with $T=0.005T_c$, $\vec h = \Delta_0\hat{T}$, $\zeta=3$, for a straight ($\kappa=0$) and semi-circular ($\kappa L_F=\pi$) ferromagnetic wire. (a) $L_F=2\xi_S$ Increasing the curvature causes the triplet contribution to change sign. (b) $L_F=6\xi_S$ Increasing the curvature causes the singlet contribution to be neglegible with respect to the triplet one, signaling that the charge current is transported almost exclusively by the triplet correlations.}
	\label{IQ_st}	
\end{figure}

Two interesting effects of the curvature appear immediately from our results. First, we show in Fig.~\ref{IQc_L2} that it is possible to induce a $0-\pi$ transition in the junction by changing the curvature of the ferromagnetic wire while keeping its length fixed. Secondly, we will show in Fig.~\ref{IQ_st} that even for a long junction, where the singlet contribution to the supercurrent is negligible, a Josephson effect still appears for a non-zero $\kappa$ due to the presence of long ranged triplets.

In Fig.~\ref{IQc_L2} we plot the absolute value of the critical current as given by \Eq\eqref{Q_current} as a function of the curvature $\kappa$ of the ferromagnet across the junction for different lengths $L_F$. From the figure we see that starting in the $0$ state with a straight wire, increasing the curvature results in a decreasing magnitude of the critical current, until it completely disappears for a certain $\kappa$, indicating a $0-\pi$ transition. A further increase in the curvature produces a revival of the critical current, which now flows in the opposite direction with respect to the straight case. We also note that increasing the length of the ferromagnet reduces not only the overall magnitude of the critical current but also the curvature at which the $0-\pi$ transition takes place.

In order to better understand how this $0-\pi$ transition appears, and to show that the role of the triplets is crucial in tuning it, we split the charge current into singlet and triplet contributions, $I_0$ and $I_t$ respectively.  It can be shown that the charge current given by \Eq\eqref{Q_current} only depends on the anomalous Green function $f$ which is the off-diagonal block matrix in the retarded Green function. We define $f=(f_0+\bm{d}\cdot\bm{\sigma})i\sigma_y$, with $f_0$ representing the singlet contribution and $\bm d = (d_T,d_N,d_z)$ the d-vector representing the triplet contribution, and obtain that the charge current can be written as $I_Q/I_{Q0}=I_0+I_t$, where $I_t = I_T+I_N+I_z+I_\kappa$ and:

\begin{subequations}\begin{align}
	I_0 &= -8\!\!\int_0^\infty \!\! d\varepsilon\Real\left\{\tilde{f}_0\partial_sf_0-f_0\partial_s\tilde{f}_0\right\}\tanh(\beta\varepsilon/2),\\
	I_j &= 8\!\!\int_0^\infty \!\! d\varepsilon\Real\left\{\tilde{d}_j\partial_sd_j-d_j\partial_s\tilde{d}_j\right\}\tanh(\beta\varepsilon/2),\\
	I_\kappa &= 16\kappa\!\!\int_0^\infty \!\! d\varepsilon\Real\left\{\tilde{d}_Nd_T-\tilde{d}_Td_N\right\}\tanh(\beta\varepsilon/2),
\end{align}\end{subequations}

\noindent with $j=(T,N,z)$. The terms $I_0$ and $I_j$ represent the contribution coming from the singlet and triplets with spin aligned in the $j$ direction respectively. The last term $I_\kappa$ instead defines an inverse Edelstein term due to the curvature. This kind of contribution appears whenever the d-vector undergoes a rotation and is therefore non zero only in the presence of finite curvature and/or spin-orbit coupling \cite{Amundsen2017}.

In Fig.~\ref{IQ_st}(a) we plot these different contributions to the charge current for two different values of $\kappa$ and $L_F=2\xi_S$. It can be seen that for $\kappa=0$ triplets and singlet charge currents have opposite sign, with the triplets contribution, which comes only from the short-range component $I_T$, being generally bigger than the singlet one. Interestingly however, when increasing the curvature the triplet current changes sign, i.e. starts flowing in the opposite direction, because of the appearance of the long-range component $I_N$, while the singlet contribution does not. Hence, the $0-\pi$ transition is tuned by the curvature through its effect on the triplets. Furthermore, we note from Fig.~\ref{IQ_st}(a) that in the $\pi$-phase for $\kappa=\pi/L_F$ the singlet and triplet currents have the same sign and thus flow in the same direction. Consequently, the two contributions add up, resulting in a larger critical current in the $\pi$-phase at $\kappa=\pi/L_F$ compared to the $0$-phase at $\kappa=0$.

We point out that curvature also introduces a spin current to the system, which is absent in a straight nanowire. This \emph{exchange} spin current, as it is known in the literature, is caused by the misalignment of the magnetization in the system and is non-zero even at phase differences of $\phi = 0$ and $\phi = \pi$, where there are no charge currents~\cite{Jacobsen2016,Chen2014}. The magnitude of the spin current is affected by the curvature, thereby providing means by which it can be externally manipulated.

To highlight that the triplets generate a long range Josephson effect, we consider a long junction, $L_F=6\xi_S$, and plot in Fig.~\ref{IQ_st}(b) separately singlet and triplet contributions for a straight ($\kappa=0$) and semi-circular ($\kappa L_F=\pi$) wire. We see that, while for $\kappa=0$ the triplet term is essentially zero and the singlet term is finite, for $\kappa L_F=\pi$ the singlet contribution is negligible compared to the triplet one, which additionally presents a long-range component $I_N$ dominating over the short-range one $I_T$. Going from a straight to a semi-circular ferromagnet produces a significant singlet to triplet conversion, of which component in the normal direction is long-ranged, i.e. $d_N=|\bm{d}\times\bm{\hat{h}}|$, since in the case considered the $d_z$ component is zero. In the simple example of a long wire with constant curvature chosen here, the magnitude of this LRT component is quite small, but we explain how this can be increased and manipulated below. 

To better understand the role played by the curvature, it is useful to consider the weak proximity effect, meaning that $\left|\gamma_{ij}\right|\ll1$ and $N\simeq1$. The $\gamma$ matrix can be then expressed in terms of $f$: $\gamma=f/2$. We then obtain the following linearized version of the Usadel equation:

\begin{subequations}\begin{align}
	\frac{D_F}{2}\partial_s^2f_0&=-i\varepsilon f_0-i\bm{h}\cdot\bm{d}, \label{eq:f0}\\
	\frac{D_F}{2}\!\left(\partial_s^2d_T\!-\!2\kappa\partial_sd_N\!\right)\!\!&=\!\left(\!-i\varepsilon\!+\!\frac{D_F\kappa^2}{2}\!\right)\!d_T\!-\!if_0h_T, \label{eq:dT}\\
	\frac{D_F}{2}\!\left(\partial_s^2d_N\!+\!2\kappa\partial_sd_T\!\right)\!\!&=\!\left(\!-i\varepsilon\!+\!\frac{D_F\kappa^2}{2}\!\right)\!d_N\!-\!if_0h_N, \label{eq:dN}\\
	\frac{D_F}{2}\partial_s^2d_z&=-i\varepsilon d_z-if_0h_z.
\end{align}\end{subequations}

By inspecting the linearized Usadel equation for the triplet components, given in \cref{eq:f0,eq:dT,eq:dN}, we see that the curvature produces a Dyakonov-Perel term, describing the spin-relaxation due to precession around the exchange field. A curvature of $\kappa L_F=\pi$ gives a strong spin-relaxation term which causes a fast decay even for the LRT component.  From a qualitative perspective we can see that, since the exchange field varies with the position, a LRT component flowing through the wire will acquire an increasing component parallel to $\bm{h}$, i.e. a quickly decaying short-range component. The SRT component likewise acquires a LRT, but the conversion region is restricted to the typical decay of the SRT $\sim 1/\sqrt{h}$. In order to maximise the LRT generation from the SRT, one should therefore have a region of high curvature over the spatial decay of the SRT near the superconducting interface, and then minimal or zero curvature beyond. Alternatively, one may start with an intrinsically triplet superconductor, or have a compensating spin-orbit field in the ferromagnet that can negate the effect of the curvature.

\vspace{3mm}

\textit{Concluding remarks}.---
We have shown that curvature is a designable and tunable parameter that can generate and control long-ranged supercurrents in diffusive SFS Josephson junctions without any magnetic inhomogeneities or intrinsic SOC. The system displays a curvature-controlled $0-\pi$ transition, which can be manipulated dynamically in-situ with a single sample. This can facilitate experimental investigation of the transition, and improve our understanding of the coexistence of superconductivity and magnetism in different phases. In the longer term this opens a diverse new toolkit for design and control of diffusive superconducting spintronic systems, and may be a useful implementation in solid state quantum computing. Since this field is still in its infancy, with several exciting directions still to be explored, we anticipate that curvature in such systems will be integral to the new generation of spintronic designs. 

\vspace{3mm}

\begin{acknowledgments}
We thank P. Gentile for useful discussions and H. G. Hugdal for his helplful insights. The computations have been performed on the SAGA supercomputer provided by UNINETT Sigma2 - the National Infrastructure for High Performance Computing and Data Storage in Norway. We acknowledge funding via the ``Outstanding Academic Fellows'' programme at NTNU, the Research Council of Norway Grant number 302315, as well as through its Centres of Excellence funding scheme, project number 262633, “QuSpin”.
\end{acknowledgments}

\bibliographystyle{apsrev4-2}
\bibliography{Curvbib.bib} 

\end{document}